\documentclass[showpacs,showkeys,11pt,
preprint,preprintnumbers,nofootinbib,floatfix,
groupedaddress,superscriptaddress,amsmath,amssymb]{revtex4}

\usepackage{amsfonts}
\usepackage{amsmath}
\usepackage{graphicx}
\graphicspath{{plots/}}
\usepackage[section]{placeins}
\usepackage{booktabs}
\usepackage{url}

\usepackage[colorlinks=true,linkcolor=blue,citecolor=blue,urlcolor=blue]{hyperref}
\usepackage[nameinlink,noabbrev]{cleveref}
\crefname{section}{Sec.}{Secs.}
\Crefname{section}{Sec.}{Secs.}
\crefname{subsection}{Sec.}{Secs.}
\Crefname{subsection}{Sec.}{Secs.}
\crefname{figure}{Fig.}{Figs.}
\Crefname{figure}{Figure}{Figures}
\crefname{table}{Table}{Tables}
\Crefname{table}{Table}{Tables}
\crefname{equation}{Eq.}{Eqs.}
\Crefname{equation}{Equation}{Equations}

\newcommand{\vlltau}{\tau^\prime}
\newcommand{\vllnu}{\nu^\prime}
\newcommand{\mvll}{m_{\rm VLL}}

\begin{document}
\raggedbottom

\title{Chirality Observables in Vector-Like Lepton Doublet Pair Production with Polarized Beams at a 3~TeV Lepton Collider}

\author{Haroon Sagheer}
\email{haroonsagheer663@gmail.com}
\affiliation{Riphah International University, Islamabad, Pakistan}

\author{Ijaz Ahmed}
\email{ijaz.ahmed@fuuast.edu.pk}
\affiliation{Federal Urdu University of Arts, Science and Technology, Islamabad, Pakistan}

\author{Jamil Muhammad}
\email{mjamil@konkuk.ac.kr}
\affiliation{Sang-Ho College and Department of Physics, Konkuk University, Seoul 05029, South Korea}

\date{\today}

\begin{abstract}
We study how pair production of a mass-degenerate weak-isodoublet vector-like lepton (VLL) depends on longitudinal beam polarization at the 3~TeV Compact Linear Collider (CLIC), for VLL masses of 1.1--1.4~TeV and mixing in the tau sector. The charged pair $e^+e^-\to\tau'\bar\tau'$ proceeds through photon and $Z$ exchange, the neutral pair $e^+e^-\to\nu'\bar\nu'$ through $Z$ exchange alone. For the baseline electron polarization $P_{e^-}=-0.8$, the tree-level rates at $\mvll=1.2$~TeV are 15.77~fb (charged) and 4.75~fb (neutral). For an optional positron polarization $P_{e^+}=+0.3$, the left--right configuration $(-0.8,+0.3)$ raises the two rates by factors of 1.95 and 1.38 relative to unpolarized beams. The charged--neutral rate contrast reaches $D_\sigma=0.546$ there, compared with $0.08$ for the opposite configuration, and the observed left--right asymmetries are 0.57 (charged) and 0.12 (neutral). These normalized quantities are independent of the VLL mass at Born level because both channels share the same phase-space factor. Unpolarized rates at a 3~TeV muon collider coincide with the unpolarized CLIC rates to within 0.2\%, as expected from lepton universality. A counting estimate gives a Gaussian separation of the two asymmetries of $Z_A\simeq9$ for a visible fraction of $10^{-1}$ and $Z_A\simeq2.7$--$2.9$ for $10^{-2}$ at $1~{\rm ab}^{-1}$. All results are tree-level, fixed-energy, production-level quantities; initial-state radiation, beamstrahlung, detector effects and backgrounds are not included, and no exclusion or discovery reach is claimed.
\end{abstract}

\pacs{12.60.Fr, 12.60.-i, 14.60.Hi, 14.80.-j}
\keywords{Vector-like leptons, CLIC, Beam polarization, Chirality observables}

\maketitle

%%%%%%%%%%%%%%%%%%%%%%%%%%%%%%%%%%%%%%%%%%%%%%%%%%%%%%%%%%%%%%%%%%%
\section{Introduction}
\label{sec:intro}

Vector-like leptons (VLLs) are fermions whose left- and right-handed components carry identical gauge quantum numbers, so that a gauge-invariant mass term is allowed. They are a common ingredient of extensions of the Standard Model (SM) and are constrained by electroweak precision data and lepton observables \cite{PDG2024,CynolterLendvai2008,delAguila2008,IshiwataWise2013,Falkowski2014}. Their collider phenomenology depends on the gauge representation, the mass spectrum, the mixing with SM leptons and the resulting decay channels.

At the LHC, a CMS multilepton search excludes a prompt third-generation VLL doublet below $1045~{\rm GeV}$ at 95\% confidence level for its decay benchmark \cite{CMS2022VLL}, and ATLAS reports an observed limit of $900~{\rm GeV}$ for its doublet scenario \cite{ATLAS2023VLL}; further searches address other lepton flavors and decay scenarios \cite{ATLAS2025VLL}. These limits depend on the assumed branching fractions, lifetime and event selection, so applying them to the present benchmark requires matching these quantities. We therefore treat them as indicative and choose reference masses of 1200 and $1300~{\rm GeV}$, which lie above both limits and are kinematically accessible at a 3~TeV collider.

At its $3~{\rm TeV}$ stage, the Compact Linear Collider (CLIC) \cite{CLICUpdatedBaseline2016,CLIC2018,Roloff2018CLICPhysics,Robson2018CLICAccelDetector,deBlas2018CLICNewPhysics} is well suited to electroweak pair production, and its baseline foresees polarized electron beams. Pair production of VLLs at $e^+e^-$ colliders has been studied for isosinglet leptons mixing with the tau \cite{BhattiproluMartinPierce2024}, including initial-state radiation (ISR) and beam polarization \cite{Shen2025ChinPhysC}; for charged isosinglets with boosted-boson signatures at CLIC \cite{LiLianLiu2025CLIC}; and for first-generation singlets with polarized beams and detector simulation \cite{LiuMoretti2025}. Single production of first-generation doublet VLLs at CLIC has also been examined \cite{GaoLiuMoretti2026CLIC}. Here we consider the weak-isodoublet, which contains a charged and a neutral member of equal mass. The charged pair is produced through photon and $Z$ exchange, the neutral pair through the $Z$ current alone, so the two channels respond differently to the beam helicities. We characterize this response analytically and with event-generator cross sections, using the rates, a charged--neutral rate discriminator $D_\sigma$, and the left--right asymmetry. We distinguish the CLIC baseline ($P_{e^-}=\pm0.8$, $P_{e^+}=0$) from an optional upgrade with $|P_{e^+}|=0.3$ \cite{CLIC2012Polarization}, and we give unpolarized muon-collider rates as a cross-collider benchmark. A visible-fraction requirement and a counting-statistics diagnostic show how these quantities scale with luminosity; they are neither exclusion nor discovery projections. We do not compare different electroweak representations.

The doublet model is defined in \cref{sec:framework}, the polarization setup and observables in \cref{sec:setup}, the results in \cref{sec:results}, and our conclusions in \cref{sec:conclusions}.

%%%%%%%%%%%%%%%%%%%%%%%%%%%%%%%%%%%%%%%%%%%%%%%%%%%%%%%%%%%%%%%%%%%
\section{Model and Production Processes}
\label{sec:framework}

\subsection{Vector-like lepton doublet}
\label{subsec:vll_doublet_benchmark}

We use the weak-isodoublet VLL model implemented in the UFO model \texttt{VLLD\_NLO} \cite{VLLUFOrepo}, generated with FeynRules and interfaced to \textsc{MadGraph5\_aMC@NLO} \cite{Alloul2014,Degrande2012UFO,Alwall2014MG5}. The same class of doublet has been used in earlier collider studies \cite{KumarMartin2015,BhattiproluMartin2019}. No singlet or triplet representation is considered. The new leptons form an $SU(2)_L$ doublet with hypercharge $Y=-1/2$,
\begin{equation}
L'=
\begin{pmatrix}
\vllnu\\[2pt]
\vlltau
\end{pmatrix},
\qquad
Q=T_3+Y ,
\label{eq:vll_doublet}
\end{equation}
so that $\vllnu$ ($T_3=+1/2$) is neutral and $\vlltau$ ($T_3=-1/2$) has charge $-1$. Both chiralities transform identically, which permits the vector-like mass term $\mvll\bar L'L'$. The model fixes the two masses to be equal,
\begin{equation}
M_{\vllnu}=M_{\vlltau}\equiv \mvll ,
\label{eq:vll_mass_degeneracy}
\end{equation}
which is part of the benchmark definition rather than a free parameter. After electroweak symmetry breaking, the gauge interactions relevant for pair production are
\begin{align}
\mathcal{L}_{\rm gauge}
&=
-e\,\bar{\vlltau}\gamma^\mu \vlltau A_\mu
+
\frac{g}{c_W}\left(-\frac{1}{2}+s_W^2\right)
\bar{\vlltau}\gamma^\mu \vlltau Z_\mu
+
\frac{g}{2c_W}
\bar{\vllnu}\gamma^\mu \vllnu Z_\mu
\nonumber\\
&\quad
+
\frac{g}{\sqrt{2}}
\left(
\bar{\vllnu}\gamma^\mu \vlltau W^+_\mu
+
\bar{\vlltau}\gamma^\mu \vllnu W^-_\mu
\right),
\label{eq:vll_gauge_lagrangian}
\end{align}
where $e$ and $g$ are the electromagnetic and $SU(2)_L$ couplings and $s_W$ ($c_W$) is the sine (cosine) of the weak mixing angle. The heavy-lepton gauge currents are purely vector-like.

\subsection{Pair-production amplitudes}
\label{subsec:pair_production_amplitudes}

At tree level, the charged pair $e^+e^-\to \vlltau\bar{\vlltau}$ receives photon and $Z$ exchange, whereas the neutral pair $e^+e^-\to \vllnu\bar{\vllnu}$ proceeds through the $Z$ alone. With the neutral-current couplings \cite{PDG2024}
\begin{equation}
g_L^e=-\frac{1}{2}+s_W^2,
\quad
g_R^e=s_W^2,
\quad
g_{\vlltau}=-\frac{1}{2}+s_W^2,
\quad
g_{\vllnu}=\frac{1}{2},
\label{eq:z_couplings_sec2}
\end{equation}
the schematic amplitudes are
\begin{align}
\mathcal{M}(e^+e^-\to \vlltau\bar{\vlltau})
&=
\frac{e^2 Q_e Q_{\vlltau}}{s}
\left[\bar{v}(e^+)\gamma_\mu u(e^-)\right]
\left[\bar{u}(\vlltau)\gamma^\mu v(\bar{\vlltau})\right]
\nonumber\\
&\quad+
\frac{g^2}{c_W^2}
\frac{1}{s-M_Z^2+iM_Z\Gamma_Z}
\sum_{\lambda=L,R}
 g_\lambda^e\, g_{\vlltau}
\left[\bar{v}(e^+)\gamma_\mu P_\lambda u(e^-)\right]
\left[\bar{u}(\vlltau)\gamma^\mu v(\bar{\vlltau})\right],
\label{eq:taup_amplitude_sec2}
\\[4pt]
\mathcal{M}(e^+e^-\to \vllnu\bar{\vllnu})
&=
\frac{g^2}{c_W^2}
\frac{1}{s-M_Z^2+iM_Z\Gamma_Z}
\sum_{\lambda=L,R}
 g_\lambda^e\, g_{\vllnu}
\left[\bar{v}(e^+)\gamma_\mu P_\lambda u(e^-)\right]
\left[\bar{u}(\vllnu)\gamma^\mu v(\bar{\vllnu})\right],
\label{eq:nup_amplitude_sec2}
\end{align}
with $Q_e=Q_{\vlltau}=-1$ and $P_{L,R}=(1\mp\gamma_5)/2$. The numerical cross sections are generated with the UFO model, not from these schematic expressions, but the equations explain the different polarization responses of the two channels. The initial-state chirality enters through $g_L^e$ and $g_R^e$, and the final state through the distinct $Z$ couplings of $\vlltau$ and $\vllnu$. In the charged channel the photon and $Z$ amplitudes interfere constructively for left-handed electrons ($g_L^eg_{\vlltau}>0$) and destructively for right-handed electrons ($g_R^eg_{\vlltau}<0$).

For massless initial-state leptons and a common final-state mass $M$, the angle-integrated Born cross sections of these purely vector currents share the kinematic factor
\begin{equation}
K(M,s)=\beta(3-\beta^2),
\qquad
\beta=\sqrt{1-\frac{4M^2}{s}},
\label{eq:common_pair_phase_space}
\end{equation}
so that $\sigma_X(P_{e^-},P_{e^+};s,M)=K(M,s)\,C_X(P_{e^-},P_{e^+};s)$ for $X=\vlltau\bar{\vlltau},\vllnu\bar{\vllnu}$, where $C_X$ contains the couplings, propagators and polarization weights and does not depend on $M$. At fixed $s$, cross-section ratios, $D_\sigma$ and asymmetries formed from these rates are therefore independent of $M$ at Born level. This factorization does not hold unchanged once radiation, beam-energy spread or mass-dependent selection efficiencies are included.

\subsection{Mixing, decays and generated samples}
\label{subsec:mixing_assumptions}

The UFO model contains mixing terms that let the heavy leptons decay into SM leptons and electroweak or Higgs bosons \cite{VLLUFOrepo,Falkowski2014,KumarMartin2015,BhattiproluMartin2019,BhattiproluMartinPierce2024}. We take the mixing to be in the tau sector,
\begin{equation}
\epsilon_{\tau' e_R}=\epsilon_{\tau' \mu_R}=0,
\qquad
\epsilon_{\tau' \tau_R}=0.1 .
\label{eq:vll_mixing_benchmark}
\end{equation}
Vanishing electron mixing avoids additional electron-induced contributions to pair production, so the production rates are governed by the gauge interactions of \cref{eq:vll_gauge_lagrangian}; the mixing fixes only the widths and decay patterns. Signal events were decayed with \textsc{MadSpin}: $\tau'^-\tau'^+\to(\tau^-Z,\tau^+Z)$, $(\tau^-h,\tau^+h)$, $(\tau^-Z,\tau^+h)$ and $(\tau^-h,\tau^+Z)$, and $\nu'\bar\nu'\to(\tau^-W^+,\tau^+W^-)$, with $Z\to\ell^+\ell^-$, $W\to\ell\nu$ and $h\to\tau^+\tau^-$ and stable taus in the LHE files. We generated 50\,000 events for each mass, decay topology and polarization setting at CLIC and for each mass and topology at the muon collider, together with the matching SM backgrounds ($\tau^+\tau^-ZZ$, $\tau^+\tau^-Zh$, $\tau^+\tau^-hh$, $\tau^+\tau^-W^+W^-$ and the $WW$, $ZZ$, $Zh$, $t\bar t$ controls). These samples are intended for a follow-up selection-level study; the present paper uses only the hard-process cross sections, which are quoted before decays.

\subsection{Scope}
\label{subsec:scope}

All cross sections are tree-level (the model label \texttt{NLO} notwithstanding) and are evaluated at fixed $\sqrt{s}=3~{\rm TeV}$. They omit ISR, beamstrahlung and the CLIC luminosity spectrum, which change the rates most strongly for masses close to threshold. No detector simulation, selection, background, systematic uncertainty or $CL_s$ construction is included, and no exclusion or discovery sensitivity is claimed. Every quantity below that combines rates with a luminosity or a visible fraction is a conditional production-level diagnostic.

%%%%%%%%%%%%%%%%%%%%%%%%%%%%%%%%%%%%%%%%%%%%%%%%%%%%%%%%%%%%%%%%%%%
\section{Beam Polarization and Observables}
\label{sec:setup}

\subsection{Collider and mass benchmarks}
\label{subsec:collider_mass_benchmarks}

We consider $\sqrt{s}=3~{\rm TeV}$ and $\mvll=1100$, $1200$, $1300$ and $1400~{\rm GeV}$, corresponding to $2\mvll/\sqrt{s}=0.73$, $0.80$, $0.87$ and $0.93$; pair production closes at $\mvll=1.5~{\rm TeV}$. The reference masses are $1200$ and $1300~{\rm GeV}$ (see \cref{sec:intro}). The initial states are $e^-e^+$ at CLIC and $\mu^-\mu^+$ at the muon collider. The generated muon-collider samples are unpolarized, so they support only an unpolarized rate comparison; we assign no polarized-muon asymmetries without a dedicated sample or a validated helicity reweighting.

\subsection{Polarization conventions and benchmark settings}
\label{subsec:beam_polarization_convention}

The longitudinal polarizations of the electron and positron beams are $P_{e^-}$ and $P_{e^+}$, with $P=-1\,(+1)$ denoting a fully left-handed (right-handed) beam. For negligible lepton mass, helicity and chirality labels coincide. The CLIC baseline provides $|P_{e^-}|$ up to $0.8$ with $P_{e^+}=0$; a positron polarization of up to $30\%$ has been discussed as an upgrade but is not part of the baseline \cite{CLIC2018,CLIC2012Polarization}. Settings with $P_{e^+}\neq0$ are therefore a prospective scenario and are always labelled as such. The labels LR, RL, LL and RR denote the sign pattern of $(P_{e^-},P_{e^+})$; LR means a dominantly left-handed electron beam and a dominantly right-handed positron beam. The configurations used are listed in \cref{tab:polarization_benchmarks}. If a machine stage uses other polarizations, the formulae below apply directly.

\begin{table}[!htbp]
\centering
\caption{Beam-polarization configurations. The $P_{e^+}=0$ rows belong to the CLIC baseline; all rows with $P_{e^+}\neq0$ assume the optional positron-polarization upgrade. The effective luminosity and polarization follow \cref{eq:effective_lumi_polarization}; the cross sections themselves are computed from \cref{eq:polarized_cross_section}.}
\label{tab:polarization_benchmarks}
\renewcommand{\arraystretch}{1.1}
\begin{tabular}{lcccc}
\toprule
Label & $P_{e^-}$ & $P_{e^+}$ & $\mathcal{L}_{\rm eff}/\mathcal{L}$ & $P_{\rm eff}$ \\
\midrule
Unpolarized          & $0.0$  & $0.0$  & 0.50 & $0.000$ \\
$e_L$ (baseline)     & $-0.8$ & $0.0$  & 0.50 & $-0.800$ \\
$e_R$ (baseline)     & $+0.8$ & $0.0$  & 0.50 & $+0.800$ \\
LR                   & $-0.8$ & $+0.3$ & 0.62 & $-0.887$ \\
RL                   & $+0.8$ & $-0.3$ & 0.62 & $+0.887$ \\
LL                   & $-0.8$ & $-0.3$ & 0.38 & $-0.658$ \\
RR                   & $+0.8$ & $+0.3$ & 0.38 & $+0.658$ \\
\bottomrule
\end{tabular}
\end{table}

The effective luminosity (the fraction of luminosity in opposite-helicity collisions) and the effective polarization are
\begin{equation}
\frac{\mathcal{L}_{\rm eff}}{\mathcal{L}}
=
\frac{1-P_{e^-}P_{e^+}}{2},
\qquad
P_{\rm eff}
=
\frac{P_{e^-}-P_{e^+}}{1-P_{e^-}P_{e^+}} .
\label{eq:effective_lumi_polarization}
\end{equation}
Opposite-sign polarizations increase $\mathcal{L}_{\rm eff}/\mathcal{L}$ and $|P_{\rm eff}|$, which is why LR and RL are the natural benchmarks.

\subsection{Polarized cross sections}
\label{subsec:polarized_cross_section_decomposition}

For arbitrary longitudinal polarizations, the cross section is a weighted sum of the four ideal-helicity cross sections,
\begin{align}
\sigma(P_{e^-},P_{e^+})
&=
\frac{1}{4}
\Big[
(1-P_{e^-})(1+P_{e^+})\sigma_{LR}
+
(1+P_{e^-})(1-P_{e^+})\sigma_{RL}
\nonumber\\
&\hspace{1.0cm}
+
(1-P_{e^-})(1-P_{e^+})\sigma_{LL}
+
(1+P_{e^-})(1+P_{e^+})\sigma_{RR}
\Big],
\label{eq:polarized_cross_section}
\end{align}
where the first (second) index is the electron (positron) helicity. For $s$-channel vector exchange with massless leptons, $\sigma_{LL}=\sigma_{RR}=0$. The benchmark configurations are thus partially polarized mixtures of $\sigma_{LR}$ and $\sigma_{RL}$ rather than pure helicity states. Given the generated LR and RL rates, the ideal-helicity cross sections follow from
\begin{equation}
\begin{pmatrix}\sigma(-0.8,+0.3)\\\sigma(+0.8,-0.3)\end{pmatrix}
=\begin{pmatrix}0.585&0.035\\0.035&0.585\end{pmatrix}
\begin{pmatrix}\sigma_{LR}\\\sigma_{RL}\end{pmatrix},
\label{eq:baseline_helicity_inversion}
\end{equation}
and the baseline rates are $\sigma(-0.8,0)=0.45\,\sigma_{LR}+0.05\,\sigma_{RL}$ and $\sigma(+0.8,0)=0.05\,\sigma_{LR}+0.45\,\sigma_{RL}$. The reweighted $\sigma(-0.8,0)$ reproduces the directly generated baseline sample at the quoted precision; the $(+0.8,0)$ rate is an algebraic result, not a separate event sample. The same inversion gives the unpolarized, LL and RR entries of \cref{tab:rates}, which agree with the Born-level evaluation.

Maps over the full $(P_{e^-},P_{e^+})$ plane illustrate the mathematical dependence on the polarizations; they are not machine operating points. Only the $P_{e^+}=0$ line belongs to the CLIC baseline, and $|P_{e^+}|>0.3$ is an extrapolation of \cref{eq:polarized_cross_section}.

\subsection{Observables}
\label{subsec:analysis_observables}

\paragraph*{Left--right asymmetry.} For a process $X$ we define the observed asymmetry from the opposite-sign polarization pair,
\begin{equation}
A_{LR}^{\rm obs}(X)
=
\frac{
\sigma_X(-|P_{e^-}|,+|P_{e^+}|)-
\sigma_X(+|P_{e^-}|,-|P_{e^+}|)
}{
\sigma_X(-|P_{e^-}|,+|P_{e^+}|)+
\sigma_X(+|P_{e^-}|,-|P_{e^+}|)
},
\label{eq:alr_obs_definition}
\end{equation}
and the charged--neutral separation $|\Delta A_{LR}^{\rm obs}|=|A_{LR}^{\rm obs}(\vlltau\bar{\vlltau})-A_{LR}^{\rm obs}(\vllnu\bar{\vllnu})|$. The superscript ``obs'' distinguishes partially polarized configurations from ideal pure-helicity cross sections; it does not imply a measurement. For $P_{e^+}=0$ the same definition compares $(-0.8,0)$ with $(+0.8,0)$.

\paragraph*{Rate discriminator.} At a fixed polarization point we compare the charged and neutral rates through
\begin{equation}
D_\sigma(P_{e^-},P_{e^+})
=
\frac{
\left|\sigma(\vlltau\bar{\vlltau})-
\sigma(\vllnu\bar{\vllnu})\right|
}{
\sigma(\vlltau\bar{\vlltau})+
\sigma(\vllnu\bar{\vllnu})
}.
\label{eq:dsigma_definition}
\end{equation}
The charged rate exceeds the neutral one in all configurations considered, so $D_\sigma$ equals the signed difference and carries no further sign information.

\paragraph*{Visible yield and statistical separation.} The visible yield of a mode $X$ is
\begin{equation}
N_X=\mathcal{L}\,\sigma_X\,f_{\rm vis},
\qquad
f_{\rm vis}=\epsilon\times{\rm BR}_{\rm vis},
\label{eq:visible_yield_sec2}
\end{equation}
where $\epsilon$ is the net selection efficiency and ${\rm BR}_{\rm vis}$ the visible branching fraction; this factorization keeps the production calculation independent of decay assumptions and selections. As a reference scale we use the visible fraction that yields three events,
\begin{equation}
f_{\rm vis}^{95}=\frac{N_{95}}{\mathcal{L}\,\sigma_{\rm prod}},
\qquad
N_{95}=3 .
\label{eq:fvis95_definition_sec2}
\end{equation}
Here $N_{95}=-\ln(0.05)\simeq2.996$ is the one-sided $95\%$ upper limit on a Poisson mean for zero observed events and zero expected background \cite{PDG2024,Junk1999CLs,Cowan2011,Read2002}. It is a counting reference, not an expected limit or a detector-level exclusion.

To quantify how well the two channels can be separated by their asymmetries, we define
\begin{equation}
Z_A
=
\frac{
\left|A_{LR}^{\rm obs}(\vlltau\bar{\vlltau})-A_{LR}^{\rm obs}(\vllnu\bar{\vllnu})\right|
}{
\sqrt{(\delta A_{\vlltau})^2+(\delta A_{\vllnu})^2}
},
\qquad
\delta A\simeq\sqrt{\frac{1-A^2}{N_{LR}+N_{RL}}},
\label{eq:za_definition}
\end{equation}
where the binomial counting error $\delta A$ \cite{PDG2024,Cowan2011} is evaluated separately for each channel, with $N_{LR}=\mathcal{L}_{LR}\,\sigma(-|P_{e^-}|,+|P_{e^+}|)\,f_{\rm vis}$, $N_{RL}=\mathcal{L}_{RL}\,\sigma(+|P_{e^-}|,-|P_{e^+}|)\,f_{\rm vis}$ and $\mathcal{L}_{LR}=\mathcal{L}_{RL}=\mathcal{L}_{\rm tot}/2$. $Z_A$ is a Gaussian error-propagation diagnostic, not a discovery significance.

%%%%%%%%%%%%%%%%%%%%%%%%%%%%%%%%%%%%%%%%%%%%%%%%%%%%%%%%%%%%%%%%%%%
\section{Results}
\label{sec:results}

\subsection{Production rates}
\label{sec:production_rates}

\begin{figure}[!htbp]
\centering
\includegraphics[width=0.49\textwidth]{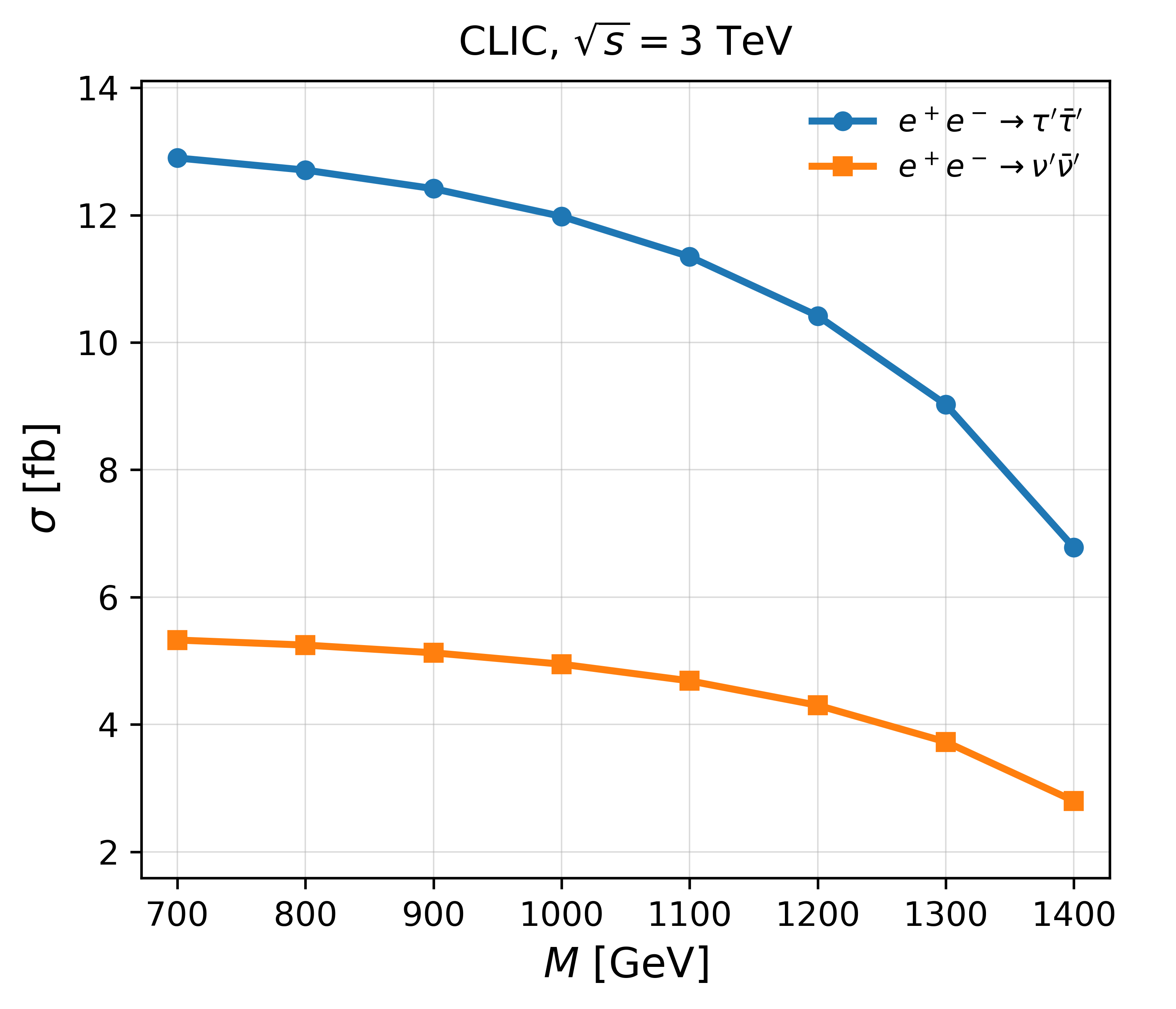}\hfill
\includegraphics[width=0.49\textwidth]{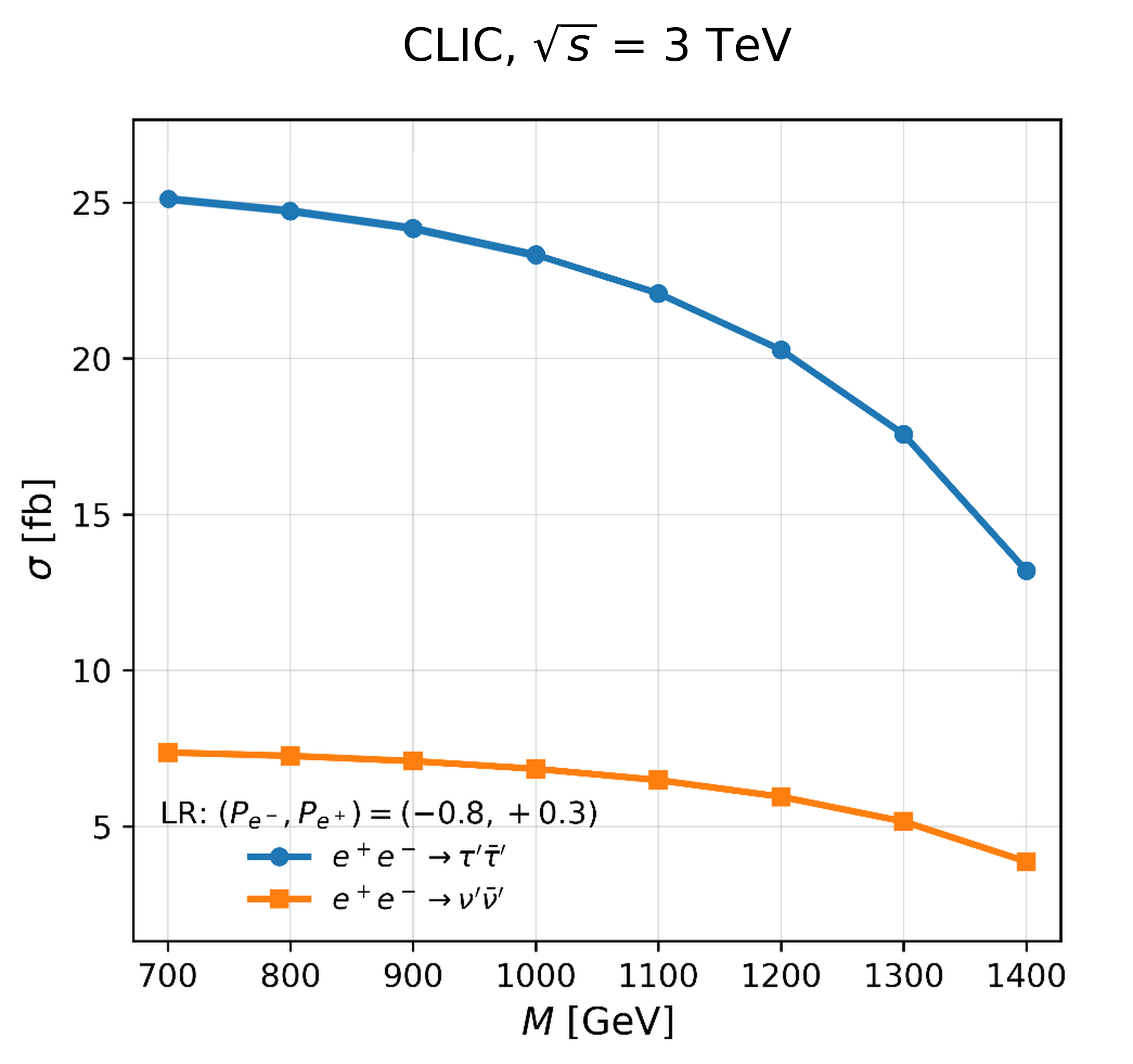}
\caption{Production cross sections versus the common VLL mass $M$ at CLIC, $\sqrt{s}=3~{\rm TeV}$, for unpolarized beams (left) and the LR configuration $(P_{e^-},P_{e^+})=(-0.8,+0.3)$ (right). The decrease with mass reflects the shrinking phase space; the larger charged rate reflects the additional photon exchange. The LR configuration enhances the charged channel more strongly.}
\label{fig:mass_scans}
\end{figure}

\Cref{fig:mass_scans} shows the mass dependence of the two rates for unpolarized beams and for the LR configuration. Both fall with $\mvll$ as phase space closes toward the $1.5~{\rm TeV}$ threshold, and the charged rate is the larger over the whole range because of the additional photon exchange. The LR configuration raises both rates, as listed in \cref{tab:rates}: by a factor $R_{\rm LR}=\sigma_{\rm LR}/\sigma_{\rm unpol}=1.95$ for the charged channel and $1.38$ for the neutral channel. The different enhancement factors are the first sign that polarization separates the two channels rather than merely increasing the yield.

\begin{table}[!htbp]
\centering
\footnotesize
\setlength{\tabcolsep}{4pt}
\renewcommand{\arraystretch}{1.1}
\caption{Born-level mass scan for the LR configuration $(-0.8,+0.3)$ at $\sqrt{s}=3~{\rm TeV}$. Only $1100$--$1400~{\rm GeV}$ coincide with generated-sample masses; $700$--$1000~{\rm GeV}$ show the trend only and lie partly below the LHC limits quoted in \cref{sec:intro}. The $1200$ and $1300~{\rm GeV}$ entries agree with the generated LR rates of \cref{tab:rates} to within $0.01~{\rm fb}$. The constancy of $D_\sigma$ (variation at the $10^{-4}$ level) confirms the factorization of \cref{eq:common_pair_phase_space}.}
\label{tab:massscan}
\begin{tabular}{lcccccccc}
\toprule
$M$ [GeV] & 700 & 800 & 900 & 1000 & 1100 & 1200 & 1300 & 1400 \\
\midrule
$\sigma_{\tau'}$ [fb] & 25.11 & 24.73 & 24.17 & 23.32 & 22.09 & 20.28 & 17.57 & 13.20 \\
$\sigma_{\nu'}$ [fb]  & 7.370 & 7.262 & 7.095 & 6.847 & 6.485 & 5.953 & 5.158 & 3.874 \\
$D_\sigma$ & 0.5462 & 0.5460 & 0.5461 & 0.5461 & 0.5460 & 0.5461 & 0.5461 & 0.5461 \\
\bottomrule
\end{tabular}
\end{table}

\begin{table}[!htbp]
\centering
\footnotesize
\setlength{\tabcolsep}{4pt}
\renewcommand{\arraystretch}{1.1}
\caption{Hard-process cross sections (fb, before decays) at $\sqrt{s}=3~{\rm TeV}$. Source: G = generated sample; R = algebraic helicity reweighting of the generated LR/RL rates (\cref{eq:baseline_helicity_inversion}); B = Born-level evaluation, which for the unpolarized, LL and RR rows coincides with the same reweighting. The ratio $\sigma_{\tau'}/\sigma_{\nu'}$ and $D_\sigma$ are mass independent at Born level and are quoted at $1200~{\rm GeV}$; they differ by at most $0.001$ at $1300~{\rm GeV}$. Monte Carlo integration errors are omitted.}
\label{tab:rates}
\begin{tabular}{llccccccc}
\toprule
 & & \multicolumn{2}{c}{$\mvll=1200$~GeV} & \multicolumn{2}{c}{$\mvll=1300$~GeV} & & & \\
\cmidrule(lr){3-4}\cmidrule(lr){5-6}
Setting & $(P_{e^-},P_{e^+})$ & $\sigma_{\tau'}$ & $\sigma_{\nu'}$ & $\sigma_{\tau'}$ & $\sigma_{\nu'}$ & $\sigma_{\tau'}/\sigma_{\nu'}$ & $D_\sigma$ & Source \\
\midrule
CLIC unpolarized  & $(0,0)$        & 10.42 & 4.30  & 9.03  & 3.73  & 2.42 & 0.416 & B \\
Muon collider unpol. & $(0,0)$     & 10.41 & 4.30  & 9.03  & 3.72  & 2.42 & 0.415 & G \\
CLIC $e_L$        & $(-0.8,0)$     & 15.77 & 4.75  & 13.66 & 4.11  & 3.32 & 0.537 & G \\
CLIC $e_R$        & $(+0.8,0)$     & 5.07  & 3.85  & 4.39  & 3.33  & 1.32 & 0.137 & R \\
CLIC LR           & $(-0.8,+0.3)$  & 20.27 & 5.95  & 17.57 & 5.16  & 3.41 & 0.546 & G \\
CLIC RL           & $(+0.8,-0.3)$  & 5.557 & 4.711 & 4.816 & 4.082  & 1.18 & 0.082 & G \\
CLIC LL           & $(-0.8,-0.3)$  & 11.26 & 3.55  & 9.76  & 3.08  & 3.17 & 0.521 & B \\
CLIC RR           & $(+0.8,+0.3)$  & 4.57  & 2.99  & 3.96  & 2.59  & 1.53 & 0.210 & B \\
\bottomrule
\end{tabular}
\end{table}

The unpolarized muon-collider rates agree with the unpolarized CLIC Born values to within about $0.2\%$, as expected: for massless leptons both colliders involve identical couplings. Differences between the machines therefore arise only from radiative and beam effects and from the available polarization, none of which are treated here. The baseline $e_L$ configuration already gives a charged-to-neutral ratio of $3.3$, close to the LR value of $3.4$, whereas the RL configuration brings the two rates much closer, $\sigma_{\tau'}/\sigma_{\nu'}=1.18$.

\begin{figure}[!htbp]
\centering
\includegraphics[width=0.98\textwidth]{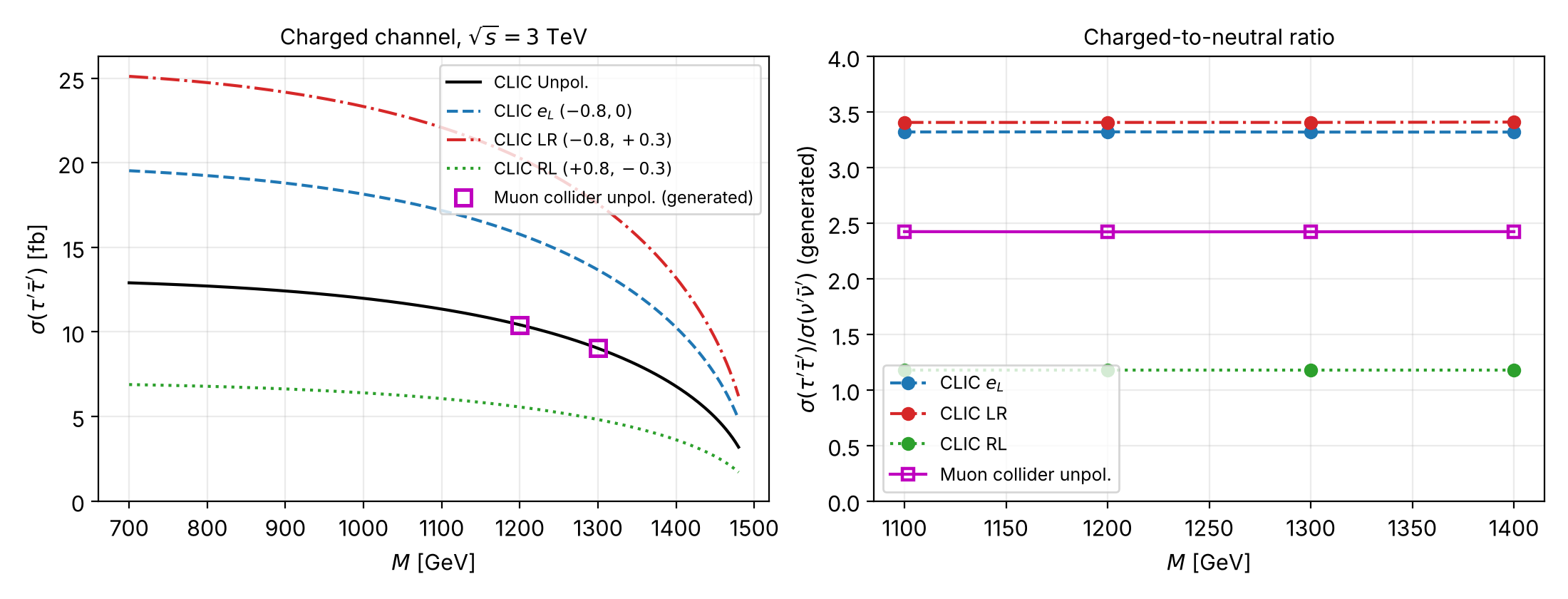}
\caption{Left: charged-channel cross section versus mass for the CLIC settings (Born-level curves) and the generated unpolarized muon-collider points at 1200 and $1300~{\rm GeV}$. Right: generated charged-to-neutral rate ratio versus mass for the three CLIC settings and the unpolarized muon collider; the ratio is flat in mass, as required by \cref{eq:common_pair_phase_space}.}
\label{fig:muon_clic}
\end{figure}

\Cref{fig:muon_clic} compares the two colliders directly. The generated unpolarized muon-collider points lie on the unpolarized CLIC curve, and the generated charged-to-neutral ratio is flat in mass for every setting: $2.42$ for the unpolarized muon collider, $3.32$ for the $e_L$ baseline, $3.41$ for LR and $1.18$ for RL. The ratio therefore depends on the beam helicities but not on the collider type or the mass. At a muon collider the unpolarized ratio of $2.42$ would change only if muon polarization were available, which would need dedicated polarized samples.

\subsection{Cross sections over the polarization plane}
\label{subsec:cross_section_polarization_maps}

\begin{figure}[!htbp]
\centering
\includegraphics[width=0.95\textwidth]{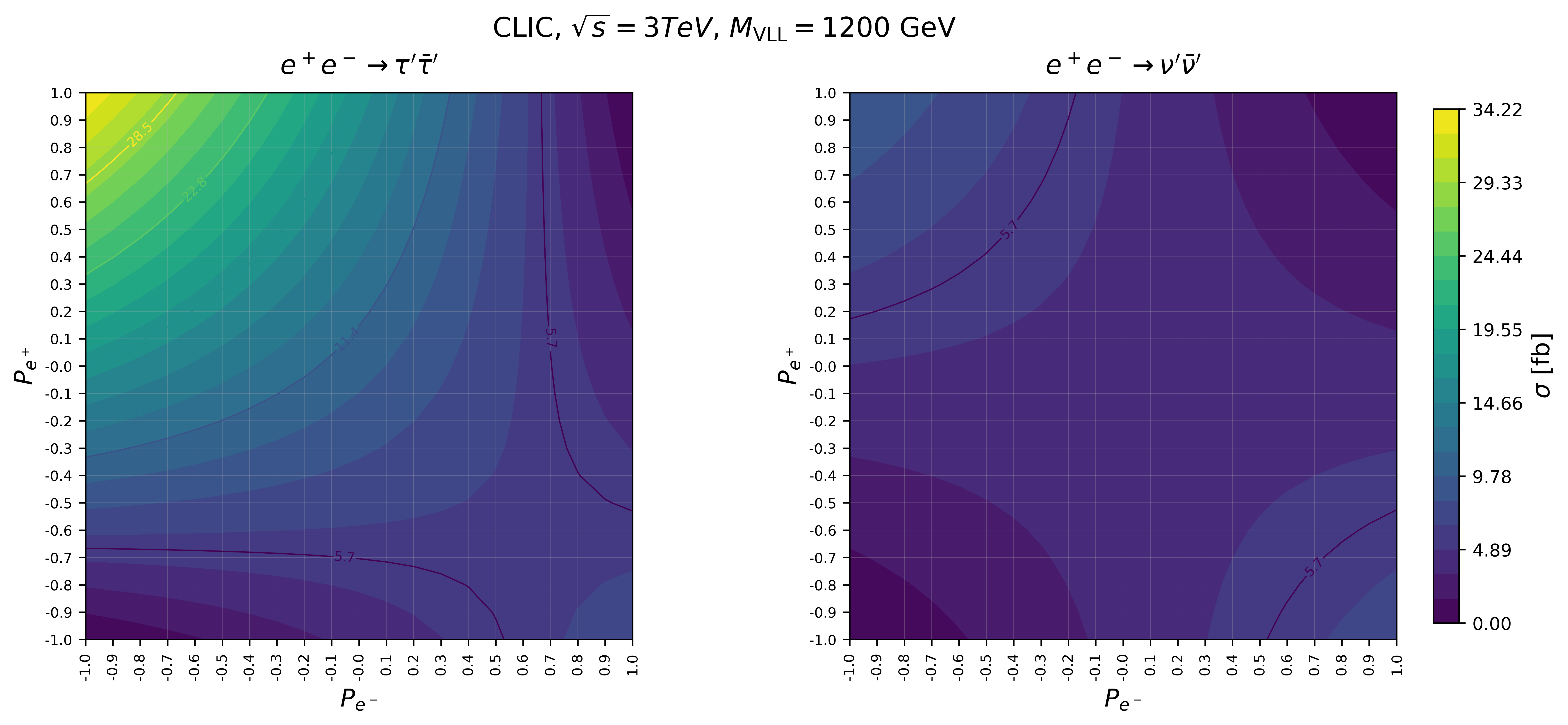}
\caption{Production cross-section contours at CLIC, $\sqrt{s}=3~{\rm TeV}$, for $\mvll=1200~{\rm GeV}$: $e^+e^-\to\vlltau\bar{\vlltau}$ (left) and $e^+e^-\to\vllnu\bar{\vllnu}$ (right), with $\sigma$ in fb. The full plane is a theoretical extrapolation, not a set of machine settings.}
\label{fig:xsec_clic_m1200}
\end{figure}

\Cref{fig:xsec_clic_m1200} shows the cross sections over the $(P_{e^-},P_{e^+})$ plane for $\mvll=1200~{\rm GeV}$. Both channels are enhanced toward the LR-rich corner; the shape is the same at all masses because of the factorization in \cref{eq:common_pair_phase_space}, while the absolute rate falls with mass. The pure same-helicity corners have vanishing rates for $s$-channel production with massless initial leptons; the benchmark points are partially polarized and have finite rates.

\subsection{Charged--neutral rate discrimination}
\label{subsec:charged_neutral_rate_discrimination}

\begin{figure}[!htbp]
\centering
\includegraphics[width=0.95\textwidth]{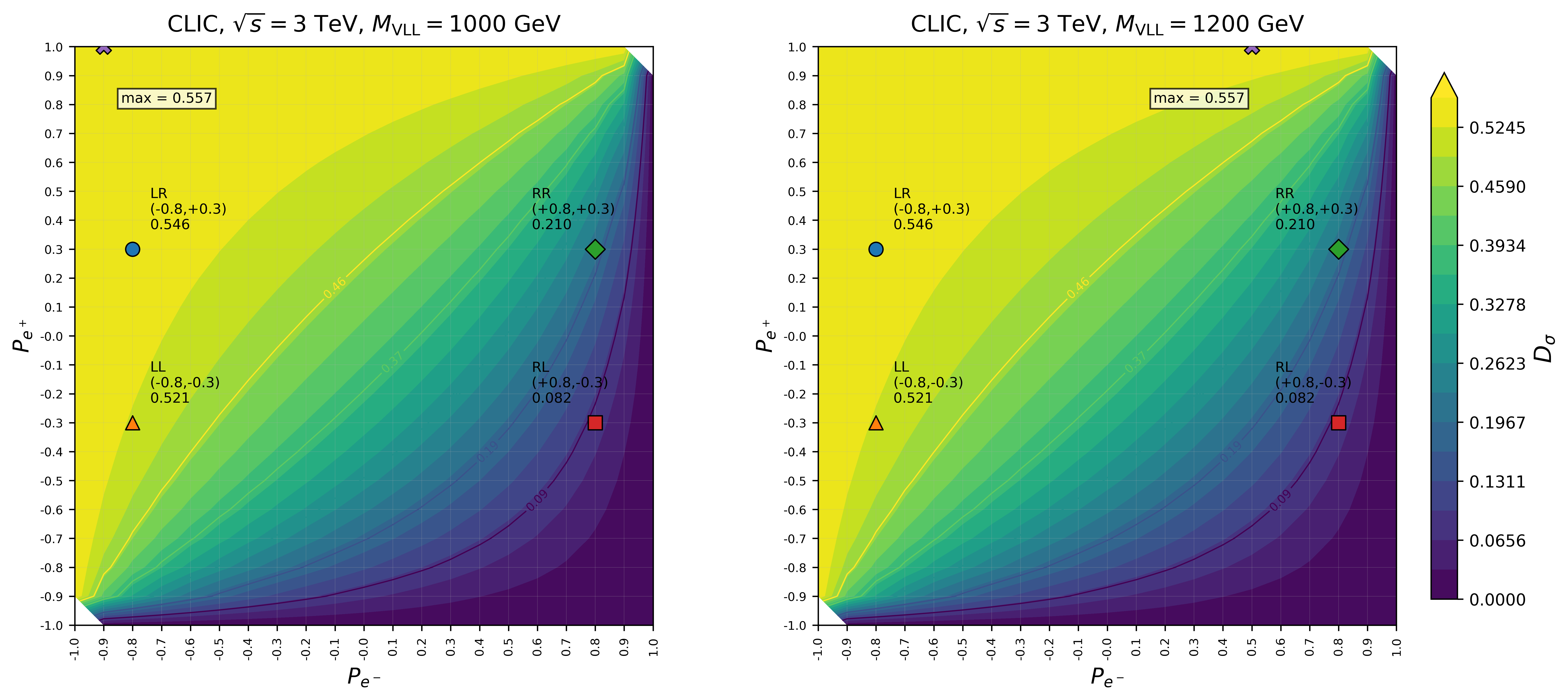}
\caption{Absolute charged--neutral rate discriminator $D_\sigma$ at CLIC, $\sqrt{s}=3~{\rm TeV}$, for $M=1000$ (left) and $1200~{\rm GeV}$ (right); the two panels coincide because $D_\sigma$ is mass independent at Born level. The marked points show the LR, LL, RR and RL benchmarks; only the $P_{e^+}=0$ line belongs to the CLIC baseline. The same-helicity corners, where the rate vanishes, are not used.}
\label{fig:dsigma_clic}
\end{figure}

\Cref{fig:dsigma_clic} shows $D_\sigma$ over the polarization plane. It measures the relative charged--neutral rate contrast at a fixed polarization point and is not an absolute yield. Its mass independence follows from \cref{eq:common_pair_phase_space}; a Born-level scan over $\mvll=700$--$1400~{\rm GeV}$ gives $D_\sigma=0.5460$--$0.5462$ at the LR point. The contrast is largest near the LR side of the plane, where $D_\sigma^{\rm LR}=0.546$ (\cref{tab:rates}). The LL point also discriminates well ($0.521$) but has a smaller effective luminosity ($0.38$ versus $0.62$ for LR), so a large contrast alone does not make a configuration attractive. The contrast is weakest on the RL side: $D_\sigma^{\rm RL}=0.082$ and, in the baseline, $D_\sigma=0.137$ for $e_R$ against $0.537$ for $e_L$. This reflects the destructive photon--$Z$ interference in the charged channel for right-handed electrons (\cref{subsec:pair_production_amplitudes}): the two channels become similar when the initial state emphasizes the right-handed electron coupling.

$D_\sigma$ compares the two channels at the same collider, mass and polarization, and thereby isolates a structural property of the doublet: photon plus $Z$ exchange for the charged state, $Z$ exchange only for the neutral state. An experimental extraction would still require the channel-dependent efficiencies and backgrounds of the two decay topologies.

\subsection{Asymmetry separation and statistical reach}
\label{subsec:asymmetry_statistical_reach}

\begin{figure}[!htbp]
\centering
\includegraphics[width=0.48\textwidth]{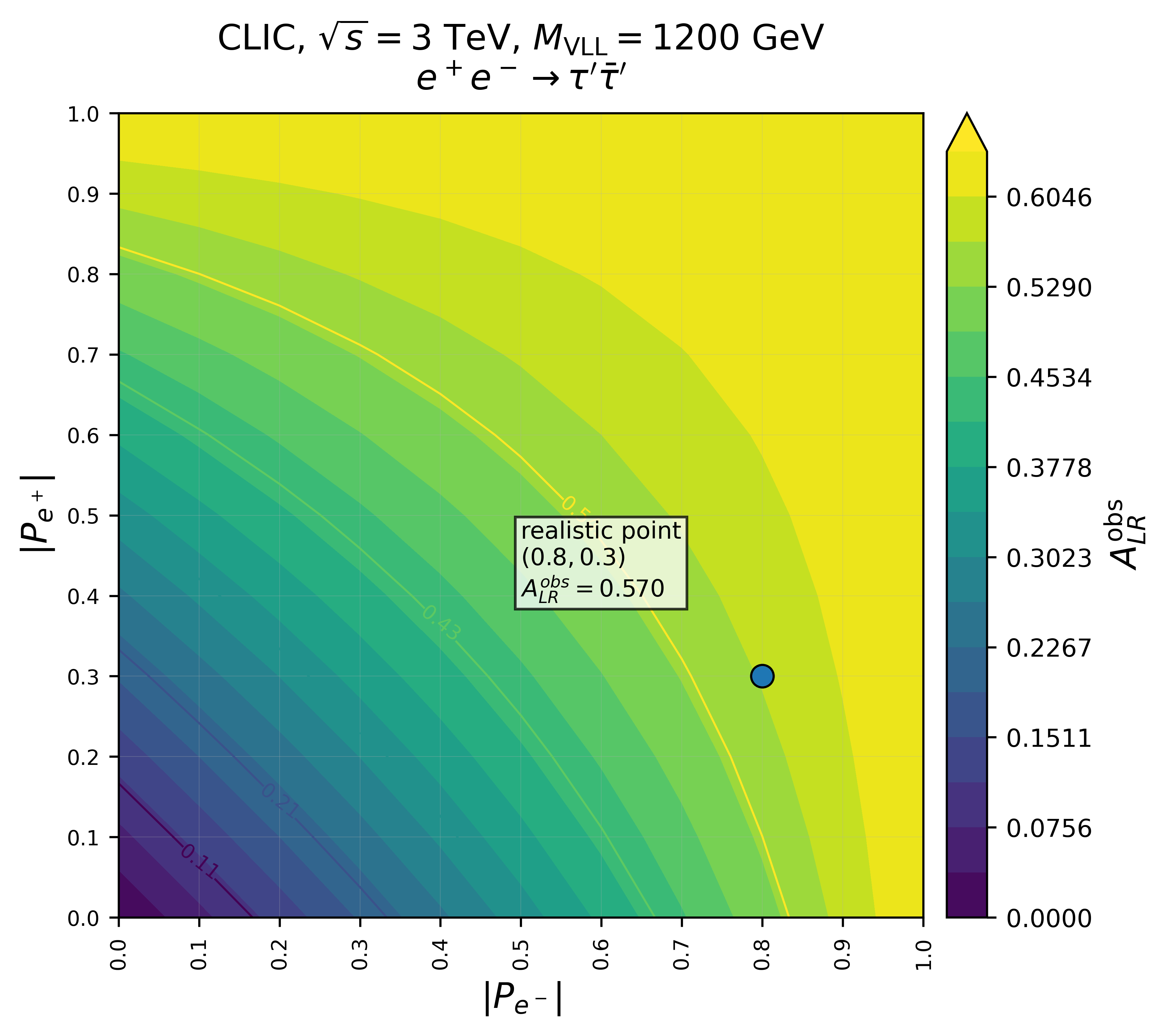}\hfill
\includegraphics[width=0.48\textwidth]{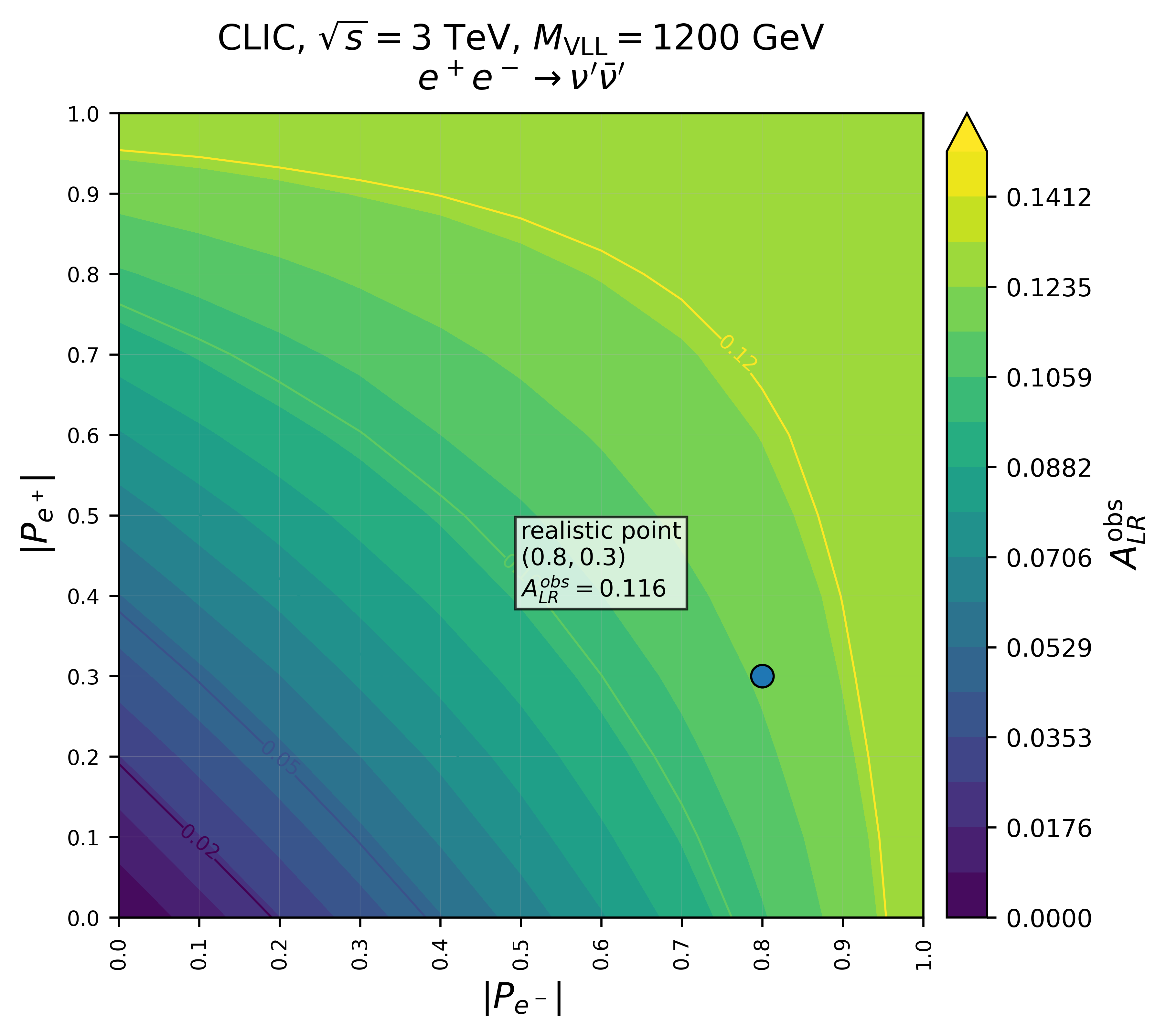}
\caption{Observed left--right asymmetry $A_{LR}^{\rm obs}$ at CLIC, $\sqrt{s}=3~{\rm TeV}$, $\mvll=1200~{\rm GeV}$, as a function of $|P_{e^-}|$ and $|P_{e^+}|$: charged channel (left) and neutral channel (right). Positive values mean that the LR-like configuration has the larger cross section. The plane is a theoretical study, not a list of machine settings; the asymmetry is mass independent at Born level.}
\label{fig:alr_maps}
\end{figure}

The asymmetry $A_{LR}^{\rm obs}$ compares two polarization settings for each channel separately and probes the chiral structure of the initial-state current together with the final-state couplings. At both reference masses (\cref{tab:rates}), $A_{LR}^{\rm obs}(\vlltau\bar{\vlltau})\simeq0.570$ and $A_{LR}^{\rm obs}(\vllnu\bar{\vllnu})\simeq0.116$, so that $|\Delta A_{LR}^{\rm obs}|\simeq0.45$; the common phase-space factor cancels in these normalized quantities. The charged state thus shows a much larger LR--RL contrast than the neutral state. For the baseline pair $(\mp0.8,0)$ the separation is smaller, $|\Delta A_{LR}^{\rm obs}|\simeq0.41$; the two numbers must not be interchanged when quoting prospective reach. \Cref{fig:alr_maps} shows the dependence of $A_{LR}^{\rm obs}$ on $|P_{e^-}|$ and $|P_{e^+}|$; the benchmark point $(0.8,0.3)$ lies on a smooth response surface.

\Cref{tab:za} gives $Z_A$ for $f_{\rm vis}=1$, $0.1$ and $0.01$ at $\mathcal{L}_{\rm tot}=100$, $500$ and $1000~{\rm fb}^{-1}$, with equal exposure in the LR and RL settings. For fixed asymmetries and cross sections, $Z_A\propto\sqrt{\mathcal{L}_{\rm tot}f_{\rm vis}}$. For $f_{\rm vis}=1$, $Z_A\simeq9$ at $100~{\rm fb}^{-1}$ and $27$--$29$ at $1000~{\rm fb}^{-1}$; each tenfold reduction of $f_{\rm vis}$ lowers $Z_A$ by $\sqrt{10}$, giving $Z_A\simeq2.7$--$2.9$ for $f_{\rm vis}=0.01$ at $1000~{\rm fb}^{-1}$.

\begin{figure}[!htbp]
\centering
\includegraphics[width=0.55\textwidth]{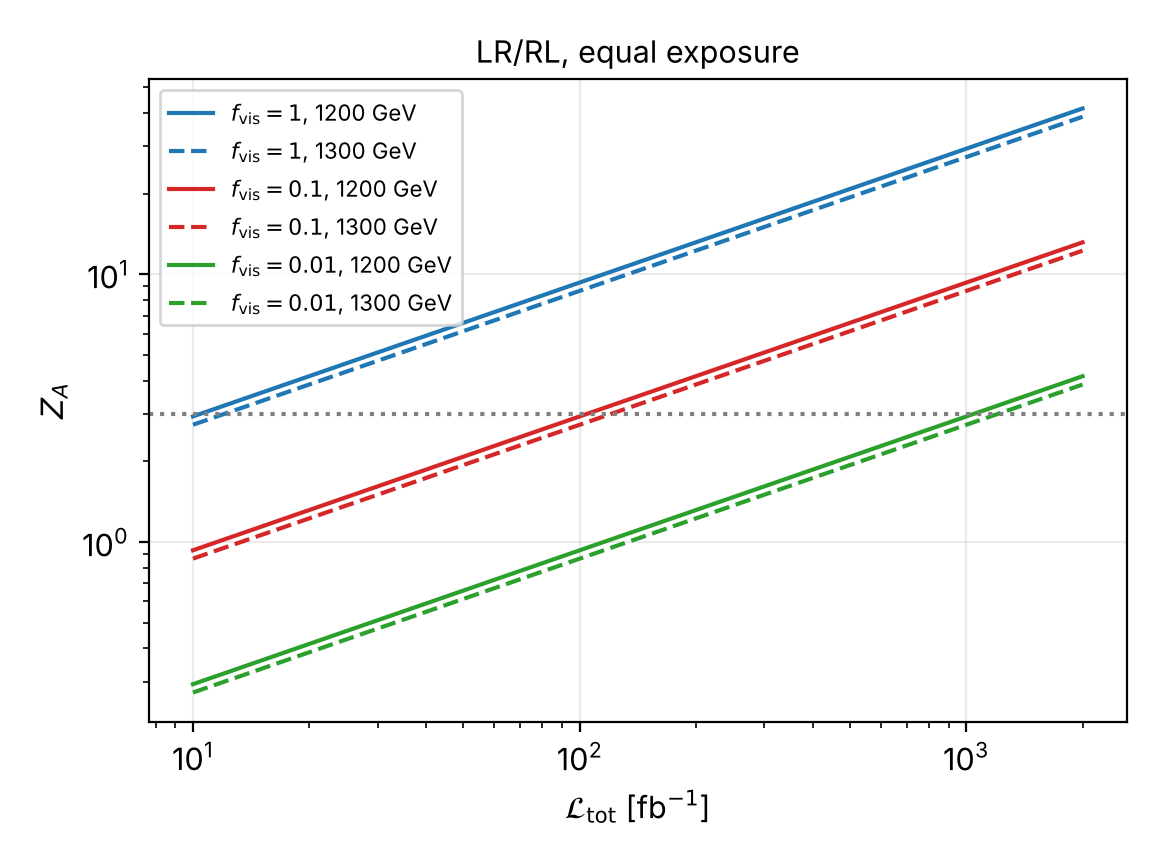}
\caption{Statistical separation $Z_A$ of the charged and neutral asymmetries versus total luminosity for $f_{\rm vis}=1$, $0.1$ and $0.01$ at $1200$ (solid) and $1300~{\rm GeV}$ (dashed), with equal LR and RL exposure. The dotted line marks $Z_A=3$. Counting statistics only.}
\label{fig:za_lumi}
\end{figure}

\begin{table}[!htbp]
\centering
\footnotesize
\renewcommand{\arraystretch}{1.1}
\caption{Statistical separation $Z_A$ of the charged and neutral asymmetries for $\mathcal{L}_{\rm tot}=100$, $500$ and $1000~{\rm fb}^{-1}$ (equal LR and RL exposure). The values include counting statistics only and use the same $f_{\rm vis}$ for both channels. At $1200$ ($1300$)~GeV, $A_{LR}^{\tau'}=0.570$, $A_{LR}^{\nu'}=0.116$ and $|\Delta A_{LR}|=0.453$ ($0.454$).}
\label{tab:za}
\begin{tabular}{ccrrr}
\toprule
$M$ [GeV] & $f_{\rm vis}$ & $Z_A(100)$ & $Z_A(500)$ & $Z_A(1000)$ \\
\midrule
1200 & 1    & 9.31 & 20.81 & 29.43 \\
1200 & 0.1  & 2.94 & 6.58  & 9.31  \\
1200 & 0.01 & 0.93 & 2.08  & 2.94  \\
\addlinespace
1300 & 1    & 8.67 & 19.38 & 27.40 \\
1300 & 0.1  & 2.74 & 6.13  & 8.67  \\
1300 & 0.01 & 0.87 & 1.94  & 2.74  \\
\bottomrule
\end{tabular}
\end{table}

\begin{figure}[!htbp]
\centering
\includegraphics[width=0.85\textwidth]{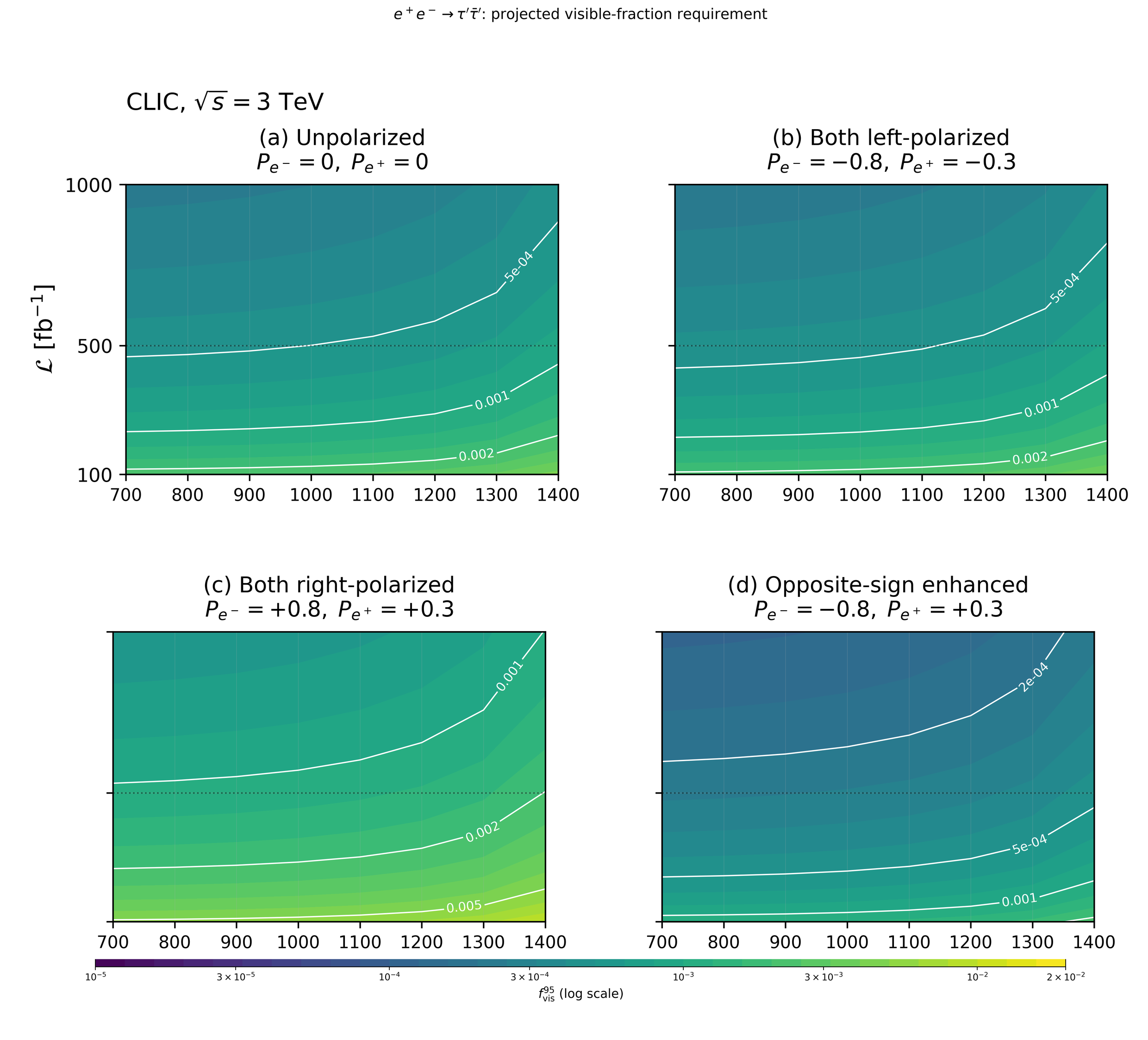}
\caption{Visible-fraction requirement $f_{\rm vis}^{95}=3/(\mathcal{L}\sigma)$ for $e^+e^-\to\tau'\bar\tau'$ at CLIC, $\sqrt{s}=3~{\rm TeV}$, for four polarization settings (the LR and RL panels assume the optional positron polarization). The dotted line marks $500~{\rm fb}^{-1}$. Lower values mean that a smaller visible branching fraction times efficiency suffices for three expected events.}
\label{fig:fvis_taup}
\end{figure}

\begin{figure}[!htbp]
\centering
\includegraphics[width=0.85\textwidth]{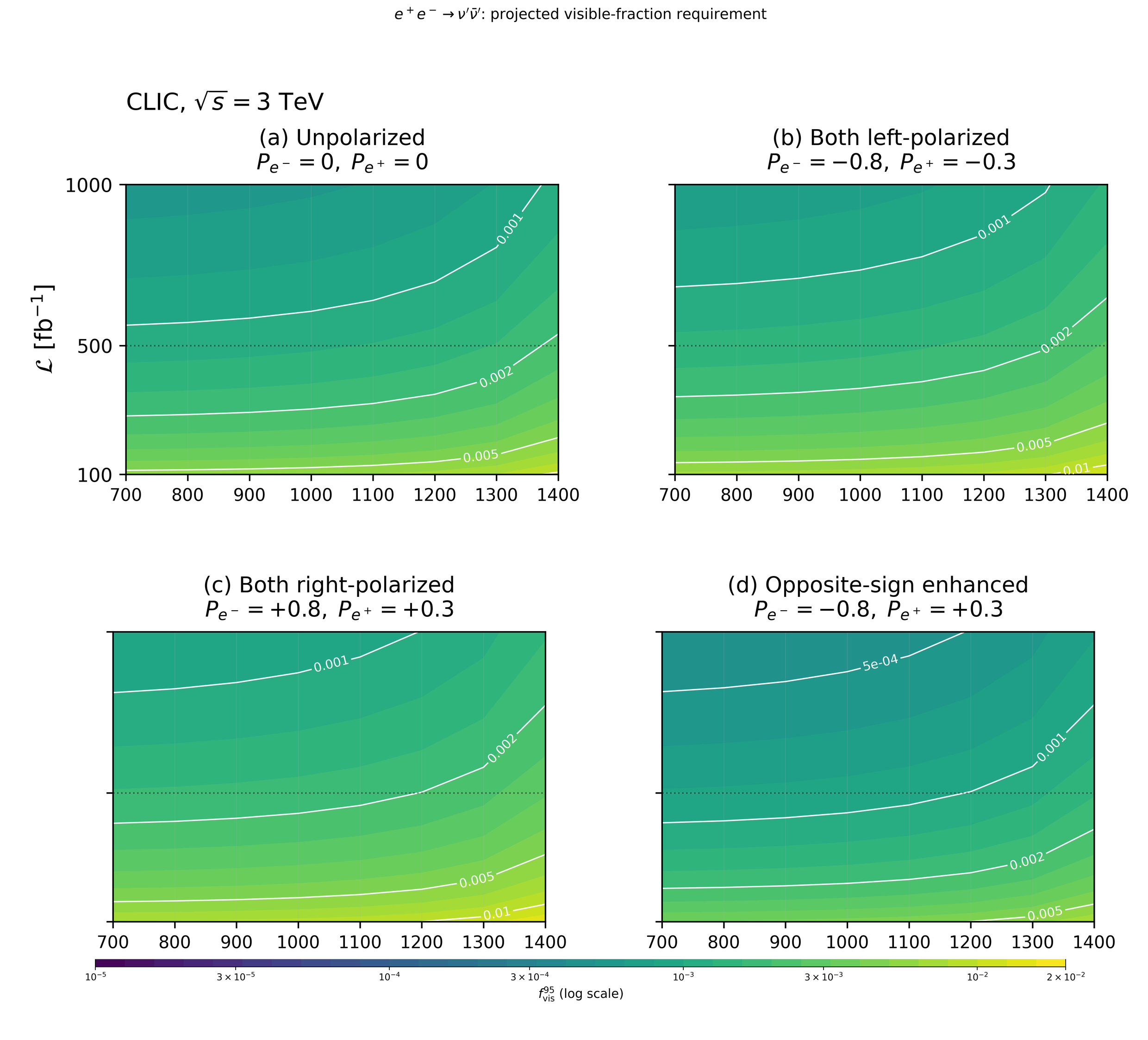}
\caption{Same as \cref{fig:fvis_taup} for $e^+e^-\to\nu'\bar\nu'$. The smaller neutral rate raises the required visible fraction.}
\label{fig:fvis_nup}
\end{figure}

The visible-fraction reference of \cref{eq:fvis95_definition_sec2} scales as $1/(\mathcal{L}\sigma)$. For the LR configuration at $1200~{\rm GeV}$ and $1~{\rm ab}^{-1}$ it is $f_{\rm vis}^{95}=1.5\times10^{-4}$ for the charged and $5.0\times10^{-4}$ for the neutral channel; it rises with mass as the cross sections fall, and it is lowest for LR among the benchmark settings. \Cref{fig:fvis_taup,fig:fvis_nup} show $f_{\rm vis}^{95}$ as a function of mass and luminosity for the charged and neutral channels. The required visible fraction falls as $1/\mathcal{L}$ and rises toward the kinematic threshold, and the neutral channel requires a visible fraction about $3.4$ times larger than the charged channel in the LR configuration. It is smallest for LR in both channels; the largest values occur for RL in the charged channel and for unpolarized beams in the neutral channel, as the rates of \cref{tab:rates} dictate. Because $f_{\rm vis}$ combines selection efficiency and visible branching fraction, a value such as $10^{-3}$ can arise from a small branching fraction, a tight selection or both. These maps are production-level scales against which future decay-specific analyses can be compared, not exclusion regions.

Several caveats apply. $Z_A$ contains statistical uncertainties only and assumes the same $f_{\rm vis}$ in both channels. The Gaussian approximation fails at small event counts, where a Poisson likelihood treatment is needed. The equal LR and RL exposure differs from the conventional CLIC run plan, which allocates about $80{:}20$ between the two electron-polarization signs \cite{CLIC2018}; the $Z_A$ values are therefore controlled illustrations and not predictions for the baseline programme. They are not discovery or exclusion contours.

%%%%%%%%%%%%%%%%%%%%%%%%%%%%%%%%%%%%%%%%%%%%%%%%%%%%%%%%%%%%%%%%%%%
\section{Conclusions}
\label{sec:conclusions}

We have studied tree-level pair production of a mass-degenerate VLL doublet at a 3~TeV lepton collider for $\mvll=1200$ and $1300~{\rm GeV}$, above the quoted CMS and ATLAS limits provided the decay and lifetime assumptions match. The charged channel receives photon and $Z$ exchange, the neutral channel $Z$ exchange only, so their polarization responses differ. For the LR configuration, the charged rates are $20.27$ and $17.57~{\rm fb}$ and the neutral rates $5.95$ and $5.16~{\rm fb}$ at the two masses, with enhancement factors of $1.95$ and $1.38$ over unpolarized beams. The charged--neutral rate contrast is $D_\sigma\simeq0.546$, and the observed asymmetries $A_{LR}^{\rm obs}\simeq0.57$ (charged) and $0.116$ (neutral) give $|\Delta A_{LR}^{\rm obs}|\simeq0.45$. These normalized quantities are mass independent at Born level because the two channels share one phase-space factor, which a Born-level scan from 700 to $1400~{\rm GeV}$ confirms for $D_\sigma$ (\cref{tab:massscan}).

The nonzero positron polarization behind these benchmarks is a prospective upgrade. For the baseline $(P_{e^-},P_{e^+})=(-0.8,0)$, the charged and neutral rates are $15.77$ and $4.75~{\rm fb}$ at $1200~{\rm GeV}$ and $13.66$ and $4.11~{\rm fb}$ at $1300~{\rm GeV}$, with $D_\sigma=0.537$ and $|\Delta A_{LR}^{\rm obs}|\simeq0.41$. Unpolarized muon-collider rates ($10.41$ and $4.30~{\rm fb}$ at $1200~{\rm GeV}$) equal the unpolarized CLIC rates to within $0.2\%$, as expected for Born-level production by massless leptons, and the generated charged-to-neutral ratio of $2.42$ is independent of mass. A counting estimate at $1~{\rm ab}^{-1}$ shows that the charged and neutral asymmetries are separable with $Z_A\simeq29$ ($27$) for $f_{\rm vis}=1$, $Z_A\simeq9$ for $f_{\rm vis}=0.1$ and $Z_A\simeq2.7$--$2.9$ for $f_{\rm vis}=10^{-2}$ at $1200$ ($1300$)~GeV. For the LR configuration at $1200~{\rm GeV}$ the three-event reference requires visible fractions of $1.5\times10^{-4}$ (charged) and $5.0\times10^{-4}$ (neutral).

These results are production-level diagnostics. ISR, beamstrahlung, the CLIC luminosity spectrum, detector response, backgrounds and systematic uncertainties remain to be included. The decayed signal and background samples described in \cref{subsec:mixing_assumptions} provide the basis for such a selection-level study, and extending the model set to singlet and triplet representations would show whether the polarization response can discriminate between VLL representations.

%%%%%%%%%%%%%%%%%%%%%%%%%%%%%%%%%%%%%%%%%%%%%%%%%%%%%%%%%%%%%%%%%%%

\end{document}